\documentclass[aps,pre,twocolumn,10pt,superscriptaddress]{revtex4-1}
\usepackage{graphicx}
\usepackage{amsmath}
\usepackage[dvipsnames]{xcolor}
\usepackage{times}
\usepackage{ulem}
\usepackage{enumitem}
\usepackage{bm}
\usepackage{amsfonts}
\usepackage[version=3]{mhchem}
\usepackage{hyperref}
\usepackage{xr}%Careful this clash with hyperref package
\usepackage{verbatim}
\usepackage{lineno}

\makeatletter
\newcommand*{\addFileDependency}[1]{% argument=file name and extension
\typeout{(#1)}% latexmk will find this if $recorder=0
\@addtofilelist{#1}
\IfFileExists{#1}{}{\typeout{No file #1.}}
}\makeatother

\newcommand*{\myexternaldocument}[1]{%
\externaldocument{#1}%
\addFileDependency{#1.tex}%
\addFileDependency{#1.aux}%
}
\myexternaldocument{supplementary}  %Careful this clash with hyperref package
\newcommand*\sref[1]{%
    S\ref{#1}}

\begin{document}
%Ignore abstract title, authors and word count  from word counting
%TC:ignore 

\title{Phyllotactic structures in radially growing spatial symmetry breaking systems.}

\author{G. Facchini}
%\thanks{These test}
\thanks{These two authors contributed equally}
%\thanks{prova}
%\email{giulio.facchini@u-paris.fr}
\affiliation{Universit\'e libre de Bruxelles (ULB), Nonlinear Physical Chemistry Unit, Brussels, Belgium}
\affiliation{Laboratoire Matière et Systèmes Complexes, CNRS, Université Paris Cité, Paris, France}
\author{M. A. Budroni}
%\email{mabudroni@uniss.it}
\thanks{These two authors contributed equally}
\affiliation{Universit\'e libre de Bruxelles (ULB), Nonlinear Physical Chemistry Unit, Brussels, Belgium}
\affiliation{Dipartimento di Scienze Chimiche, Fisiche, Matematiche e Naturali, Universit\`a di Sassari, Sassari, Italy}
\author{G. Schuszter}
%\email{schuszti@chem.u-szeged.hu}
\affiliation{Universit\'e libre de Bruxelles (ULB), Nonlinear Physical Chemistry Unit, Brussels, Belgium}
\affiliation{Department of Physical Chemistry and Materials Science, University of Szeged, Szeged, Hungary}
\author{Fabian Brau}
%\email{fabian.brau@ulb.be}
\affiliation{Universit\'e libre de Bruxelles (ULB), Nonlinear Physical Chemistry Unit, Brussels, Belgium}
\author{A. De Wit}
%\email{anne.de.wit@ulb.be}
\affiliation{Universit\'e libre de Bruxelles (ULB), Nonlinear Physical Chemistry Unit, Brussels, Belgium}

%----------------------------------------------------------------------------------------
%\GFC{
%Nature physics formatting requirements: https://www.nature.com/nphys/content
%\\
%\begin{itemize}
%    \item Main text –up to 3,000 words, excluding abstract, Methods, references and figure captions
%    \item abstract 200 words
%    \item Methods –up to 3,000 additional words, appearing online only
%    \item Figure captions should be fewer than 350 words, begin with a brief introductory sentence, and describe the meaning of all error bars.
%\end{itemize}
%}

%\textbf{Word counting (not including coloured macro comments) :}
%\detailtexcount{manuscript}     % to count words!!!! UNCOMMENT to show COUNTING

%\ADW{Commentaire de Anne}
%\FB{Commentaire de Fabian}
%\MB{Comment by Marcello}
%\GS{Comment by Gabor}
%\GF{Modifications by Giulio}
%\GFC{Giulio answer to comments}

\begin{abstract}
Phyllotactic patterns, i.e. regular arrangements of leaves or seeds around a plant stem, are fascinating examples of complex structures encountered in Nature. 
In botany, their  symmetries develop when a new primordium periodically grows in the largest gap left between the previous primordium and the apex. 
Experiments using ferrofluid droplets have also shown that phyllotactic patterns can spontaneously form when identical elements repulsing each other are periodically released at a given distance from an injection center and are advected radially at a constant speed. 
A central issue in phyllotaxis is to understand whether other self-organized mechanisms can generate such patterns. 
Here, we show that phyllotactic structures also develop in the large class of spatial symmetry-breaking systems giving spotted patterns with an intrinsic wavelength,  in the case of radial growth. We evidence this experimentally on chemical precipitation patterns, and numerically on two different models describing reaction-driven phase transitions and spatial Turing patterns, respectively.  
A generalized method for the construction of this new family of phyllotactic structures is presented, which paves the way to discover them in large classes of systems ranging from spinodal decomposition, chemical, biological  or optical Turing structures, and Liesegang patterns, to name a few.
\end{abstract}
%TC:endignore 

%\pacs{47.70.Fw; 47.15.gp; 82.40.Ck}

\date{\today}
\maketitle

\section{Introduction}
\label{Introduction} 
In botany, phyllotactic patterns identify the highly ordered self-organisation of plant organs (leaves, seeds etc.) on the stem of a unique generative spiral (the so-called {\it ontological spiral})  by a constant divergence angle, $\Phi$. Consecutive elements are then visibly connected by a number ($n+m$) of spirals ({\it parastichies}), $n$ turning one way and $m$ running in the opposite direction. Interestingly, most arrangements encountered in Nature show a divergence angle close to the golden section $\Phi=2\pi(1-\tau)$ where $\tau=(-1+\sqrt{5})/2$ is the golden mean, and $n$, $m$ are two successive numbers of the  Fibonacci series $\{F_k\}=\{1,1,2,3,5,8..\}$, in which each term is the sum of the two preceding ones \cite{Jean1994,Adler1997}. 

An important connection between such botanical patterns and physical dynamical systems has next been established in experiments showing that drops of a ferrofluid repulsing each other and drifting radially at a speed $v_0$ 
%thanks to a magnetic field 
can arrange themselves along phyllotactic patterns when released with a period $T$ on a radius $r_0$ from the center of a liquid bath \cite{Douady1992}.
Phyllotaxy was there shown to be a self-organized process due to three minimal  ingredients as  proposed by Hofmeister  \cite{Hofmeister1868}, namely: 1) radial advection at a constant speed $v_0$; 2) periodic release at a frequency $1/T$ of identical elements on a given radius $r_0$; and 3) repulsive interaction between the elements. The number $(n,m)$ of parastichies is controlled by 
%the dimensionless number $G = v_0T/r_0$,  that is the Richard's \textit{plastochrone} ratio \cite{Richards1951}
the Richard \textit{plastochrone} ratio $G = v_0T/r_0$ \cite{Richards1951},
with $(n,m)$ pairs belonging to the  Fibonacci series achieved in some limits.
Meanwhile, phyllotaxis has been shown numerically \cite{Douady1996b} to also develop when a new primordium is constrained to appear around the apex where and when there is space to do so (so-called Snow and Snow rule \cite{Snow1952}). 
%A dynamic energy landscape $E(r,\theta)$ is then defined as the superposition of multiple repulsive potentials of characteristic length $d_0$ centred on the seeds. 
%New seeds appear on a circle of radius $r_0$ centred in the apex at the angular position $\theta$ wherever $E(r_0,\theta)<E_0$, where $E_0$ is a threshold energy.
A dynamic energy landscape $E(r,\theta)$ is then defined as the superposition of multiple repulsive potentials of characteristic length $d_0$ centred on the seeds. 
New seeds appear on a circle of radius $r_0$ centred in the apex at the angular position $\theta$ wherever $E(r_0,\theta)<E_0$, where $E_0$ is a threshold energy.
In this scenario, multiple seeds appear at once which favors the selection of symmetric whorled patterns (i.e. $n=m$), and the plastochrone ratio $G$ is replaced by a pure geometric equivalent $\Gamma =d_0/r_0\propto G^2$  \cite{Iterson1907}. In a more general picture, phyllotaxis results then from  optimization of packing disks of radius $\lambda$ on a curved front of length $L$, controlled by the geometric factor $\Gamma=\lambda/L$  \cite{Gole2016}.

In this context, asking whether phyllotactic patterns can be observed outside botanical and ad-hoc physical iterative systems remains a central question.
We challenge here that question, and show both experimentally and theoretically the existence of a third family of phyllotactic patterns in spatial symmetry-breaking instabilities giving spot patterns \cite{Cross1993} in the wake of a radially diffusing or advected reaction front \cite{Brau2017}.
Specifically, we introduce self-organised spiralling patterns observed experimentally on radially advected precipitation patterns and numerically in simulations of radially growing phase separation in a Cahn--Hilliard model or reaction--diffusion Turing patterns.

%The difference with previous botany or physical phyllotactic patterns is that our new structures are stationary i.e. the concentration maxima do not move significantly in space and time once formed.

\section{Phyllotactic precipitation patterns}
\label{Experimental}

\begin{figure*}[!t]
\includegraphics[width=\textwidth]{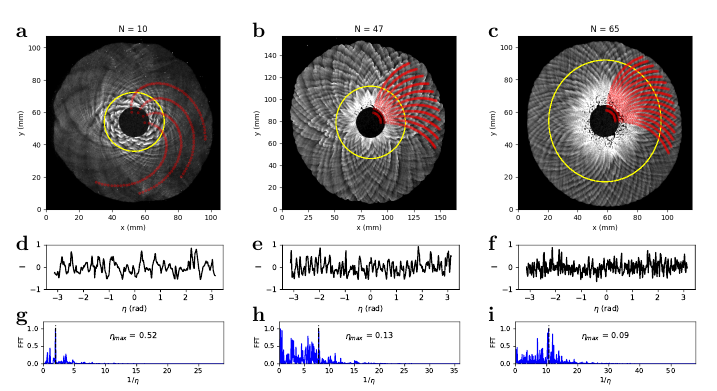}
\caption{\textbf{Experimental patterns.} \textbf{a--c:} Phyllotactic-like precipitate patterns observed when a 0.45 M aqueous solution of \ce{Na2CO3} is injected radially at a constant volumetric flow rate (\textbf{a}) $Q=0.01$ ml/min;  (\textbf{b}) $0.05$ ml/min or (\textbf{c})  $0.1$ ml/min in a Hele-Shaw cell initially filled with a 0.13 M aqueous solution of BaCl$_2$. The pH of the injected solution is adjusted to 10 by adding \ce{HCl} solution to avoid the production of barium hydroxide precipitate. 
The black central circle is masking the injection zone. The yellow circle indicates the circular path from which we extract the luminosity profile $I(\eta)$ displayed in \textbf{d--f:}. 
\textbf{g--i:} Fast Fourier transform of the profiles $I(\eta)$ shown in panels \textbf{d--f}. $\eta_{max}$ corresponds to the max of the FFT used to estimate the angle between two adjacent spirals and the number $N$ of spirals (given above each pattern). Red lines guide the eyes to recognise some of the right-turning spiral branches separated by the angle $\eta_{max}$. 
%\GS{The ''density'' of the red circles on the experimental images is too high for me and makes impossible to decide whether they overlap with the experimental spirals or they just mask them so that the reader rather belivies what's said. I propose to reduce such density similar to that in Part~a.} \GF{Working on this}.
}
\label{fig:experiments}
\end{figure*}

Precipitation patterns forming in the wake of a traveling reaction front are attracting growing interest for applications such as \ce{CO2} mineralization \cite{Schuszter2016,Schuszter2016b}, growth of self-assembled architectures  \cite{Haudin2014}, polymorph selection \cite{Bohner2014,Ziemecka2019}, microscale stamping \cite{Grzybowski2009} or biomorph growth \cite{Knoll2017}, to name a few. 
In this context, we have experimentally studied precipitation patterns in a horizontal Hele-Shaw cell (two parallel Plexiglas plates separated by a thin gap) initially filled with an aqueous solution of BaCl$_2$~\cite{Schuszter2016,Schuszter2016a,Brau2017}.
An aqueous solution of \ce{Na2CO3} is injected radially from the center of the cell at a constant flow rate. 
A white precipitate of \ce{BaCO3} is produced via an $A+B\rightarrow C $ reaction where %\GS{\sout{$A=\ce{Ba^{2+}_{~(aq)}}$, $B=\ce{CO_3^{2-}_{~(aq)}}$} 
$A=\text{CO}^{2-}_{3\,\text{(aq)}}$, $B=\text{Ba}^{2+}_\text{(aq)}$, and $C= \ce{BaCO3_{(s)}}$. 
During the dynamics, the precipitate is deposited at the outer growing rim and does not move afterwards \href{https://figshare.com/s/d61ea9dabe9d1345efe3}{(see movie S1)}.

Depending on concentrations and flow rates, spatially homogeneous \cite{Brau2017} or fingered precipitation patterns ~\cite{Schuszter2016,Schuszter2016b} can be obtained. In a given zone of the parameter space, the precipitate pattern features a phyllotactic--like structure (Fig.~\ref{fig:experiments}). 
To estimate the number $n$ and $m$ of right and left turning spirals in each pattern, we compute the luminosity $I$ of the pixels belonging to the circular path highlighted in yellow in Figs.~\ref{fig:experiments}\textbf{a--c} as a function of their angular position $\eta$ (see Figs.~\ref{fig:experiments}\textbf{d--f}).
Next, we perform a Fast Fourier transform of  $I(\eta)$ (Figs.~\ref{fig:experiments}\textbf{g--i}), and use the position of the maximum $\eta=\eta_{max}$ to estimate the angular spacing between two successive spirals. The value of $\eta_{max}$ is then used to estimate the number $N$ of spirals as $N \simeq 2\pi/\eta_{max}$ for each pattern (see top of \ref{fig:experiments}\textbf{a--c}).  
We find that in those chemical precipitation phyllotactic patterns, $n=m$ which provides highly symmetric spatial structures. This number of spirals increases with the injection flow rate and is larger than in botanical phyllotactic patterns. 

We propose that the mechanism for the formation of these spiralling precipitation patterns is due to the slaving of the precipitation process to the progression of the $A + B \rightarrow C$ front which controls the local amount of 
solid product C \cite{Antal1999}. Identifying the equivalent of seeds in plants repulsing each other to the spatial organization of precipitate domains that segregate following an energy minimisation principle allows to understand that the solid phase appears in a phyllotactic manner when associated to the radial growth.

\section{Phyllotaxy in a phase separation model}
\label{sec:cahn-hil}
Inspired by these experiments, we find that similar phyllotactic patterns can be obtained in a Cahn--Hilliard (CH) model of phase separation slaved to an $A+B\rightarrow C $ front \cite{Antal1999,Dayeh2014}. This model is a reactive version of the classical CH equation used to describe pattern formation in spinodal decomposition \cite{Cahn1958,Cahn1961}, diblock copolymer segregation \cite{Tang2005}, multiphase fluid flows \cite{Cueto-Felgueroso2014}, microstructures with elastic inhomogeneities \cite{Hu2001}, or tumor growth \cite{Ebenbeck2021}, to name a few.
Here, we integrate numerically the 2D reactive CH model in the sub-critical regime and in a radial geometry to model the precipitation of a species $C$ generated in the wake of an $A + B \rightarrow C$ front during  the radial injection of the reactant $A$ into a pool of $B$. The dimensionless model reads (see the Supplementary Information SI section \ref{ssec:CH_adim}):
\begin{subequations}
\label{cahn-hilliard}
\begin{align} 
\label{eq:A_prec}
D_t a &= \nabla^2 a - a b, \\
\label{eq:B_prec}
D_t b &= \delta_B\ \nabla^2 b - a b, \\
\label{eq:phi_prec}
\partial_t \phi &= -\nabla^2\left[\epsilon \phi - \gamma \phi^3 + \sigma \nabla^2 \phi\right] + a b.
\end{align}
\end{subequations}
where  $D_t = \partial_t + v_r\partial_r$ is the material derivative, $Q=v_r\cdot r$ is the imposed volumetric flow, 
and $\phi$ is the rescaled concentration of $C$ with $\phi=-1$ the diluted phase and $\phi=1$ the precipitate.
The coupling between the 
$A + B \rightarrow C$ 
front dynamics and precipitation is due to the source term $ab$ in Eq.(\ref{eq:phi_prec}), which takes this simple form in the limit where the reaction rate is much faster than transport (see  \ref{sec:prec_kinetic}). As seen in experiments, the precipitate is assumed not to be advected once formed. 
The coefficients $\epsilon$ and $\gamma$ determine the boundaries $\phi_m= \pm \sqrt{\epsilon/\gamma}$ and $\phi_s=\pm \sqrt{\epsilon/3\gamma}$, between the stable, metastable and unstable regions of the CH dynamics, while $\sigma$ quantifies the boundary interaction energy between different phases and controls the typical wavelength $\lambda_{CH}$ of the phase separation pattern:
\begin{equation}\label{eq:lambda_CH}
\lambda_{CH}= 2\pi\sqrt{2\sigma/(\epsilon-3\gamma\phi_0^2)},
\end{equation}	
as predicted by linear stability analysis around a spatially homogeneous solution $\phi=\phi_0$.
%as predicted by the linear stability analysis of the homogeneous reference state $\phi(t=0,x,y)=\phi_0$. 
\begin{figure*}[!thb]
\includegraphics[width= 0.9\textwidth]{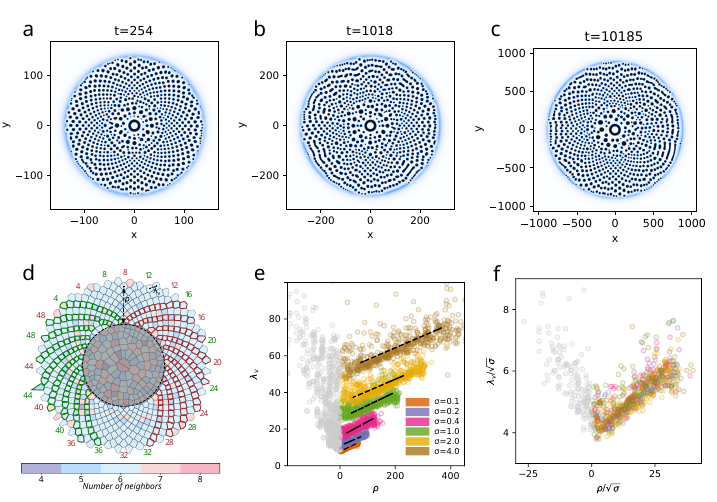}
\caption{\textbf{Numerical solutions of the Cahn--Hilliard model and pattern analysis.} \textbf{a--c:} Two-dimensional distribution of the phase field $\phi$, solution of model~(\ref{cahn-hilliard}) for  $\sigma=0.1 \ $ (\textbf{a}), $\sigma=0.4 \ $ (\textbf{b})   and $\sigma=1$ (\textbf{c})  in a regime of transverse symmetry breaking with $Q=32$, $\gamma = 1$, $\epsilon = 1$ and $\delta_B=1$~\cite{Dayeh2014}. The initial condition is $(a,b,\phi)=(0,2,-1)$ everywhere; no-flux boundary conditions are required for $\phi$ at the inlet radius $R_i=2\sqrt{2}\lambda_{CH}$, whereas $a=8, b=0$; no-flux boundary conditions for all variables at the outer radius of the system. 
Color code varies from white ($\phi=-1$, diluted phase) to blue ($\phi=1$, solid phase).
\textbf{d:} 
Voronoi diagram coloring the cells around the position of the maxima of $\phi$ depending on their number of edges, for the simulation shown in \textbf{a}. The black circle denotes the coarsening region and the corresponding cells are shaded in grey.
A selection of left (right)-turning spirals is highlighted in green (red) respectively while counting numbers are added for 1 over 4 spirals next to their ends. 48 spirals in total are counted in both directions. 
\textbf{e:} Radius $\lambda_{v}$ of the Voronoi cells as a function of their distance $\rho=r-R_0$ from the coarsening region, for simulations at variable $\sigma$. The points belonging to the coarsening region ($\rho<0$) are shaded in grey. 
\textbf{f:} Same as \textbf{e} but both axes are now rescaled by $\sqrt{\sigma}$. 
}
\label{fig:CH}
\end{figure*}

Fig.~\ref{fig:CH}\textbf{a--c} show the numerical precipitation patterns obtained for three different values of $\sigma$ with advective--diffusive radial transport.
Strikingly, the spatial self-organization of $\phi$ is reminiscent of phyllotactic patterns.
Increasing the value of $\sigma$ increases the characteristic wavelength of the pattern as shown in Fig.~\ref{fig:CH}\textbf{e} but its global shape remains self-similar, i.e. the number of parastichies remains approximately the same. 
%\GS{If the wavelength is characteristic, then how is the spiral number quazi constant? If such, then wavelength is not characteristic to the given pattern but to the entire phenomenon, and could be said that the phenomenon is not really sensitive to this paremeter.}
Similarly to the experimental structures (Figs.~\ref{fig:experiments}\textbf{a--c}), CH patterns develop radially from the center while new spots are formed along the outer expanding circular reaction front and do not move once formed \href{https://figshare.com/s/b02db9baed8f483f8f6f}{(see movie S2)}.
The radius $R$ of the circular front position scales as $R\propto t^{1/2}$ (see Fig.~\sref{sfig:CH_R_t}), which is the signature of diffusive radial transport for constant volumetric flow rate advection, i.e. $v(r)\propto 1/r$.
%\GS{It is misleading to write ,,constant flow rate advection'', since we write the same in the experimental part, but there it means constant volumetric flow rate, which in fact decays with space because of radial geometry.}. 
This indicates that the phase transition is slaved to the radial reaction front dynamics controlled by Eqs.(\ref{eq:A_prec}) and (\ref{eq:B_prec}).
Some characteristic late time coarsening  is observed near the inlet region   \cite{Chakrabarti1993}.
 
Fig.~\ref{fig:CH}\textbf{d} shows the Voronoi diagram of the pattern displayed in Fig.~\ref{fig:CH}\textbf{a}.
Outside the coarsening region, the spots are mainly arranged on an hexagonal lattice as confirmed by the histograms shown in Fig. \sref{sfig:CHneig}. This result shows that the selection of a certain spiralling mode responds to a spot packing optimisation criterion scenario \cite{Douady1996b} and, consistently, the number of left and right turning spirals is the same (see Fig.~\ref{fig:CH}\textbf{d}), i.e. the observed patterns have no chirality.

In the general picture proposed by Gol\'e et al. \cite{Gole2016}, the number of spirals is controlled by the dimensionless number $\Gamma=\lambda/L$, $L$ being the perimeter of the front where new spots appear and $\lambda$ the distance between two spots on the front.
%$\lambda$ is the distance between new spots where they appear and $L$ is the perimeter of the front where they are added.
%We stress that, similarly to the experiments (Fig.1), in the CH model $\lambda$ spots appear on peripheral circles of perimeter $L=2\pi r$.
%$\Gamma$ is then a function of the radius and is estimated by computing the local wavelength $\lambda(r)$ of the pattern, with the approximation that the pattern is axisymmetric.  
Here, spots appear on peripheral circles, thus $L=2\pi r$ while 
 $\lambda(r) \sim \lambda_{v}$  where $\lambda_{v}$ is twice the apothem of a hexagon of area $A_{v}$ and $A_{v}$ is the area of the Voronoi cells (Fig.~\ref{fig:CH}\textbf{d}). 
%the area of each Voronoi cell, and is defined as $\lambda_{v}= 2(2A_{v}/3\sqrt{3})^{1/2}$.
The histogram of Fig.\sref{sfig:CH_scale}\textbf{a} 
%\GS{labal ''a'' is missing in SI figure} 
shows the distribution of $\lambda_{v}$ for four different  values of $\sigma$. 
In Fig.~\sref{sfig:CH_scale}\textbf{b}, we observe that, if rescaled by $\sqrt{\sigma}$, all histograms collapse on each other, which indicates that $\lambda_v(\sigma)$ scales like the theoretical value $\lambda_{CH}(\sigma)$ (Eq.\ref{eq:lambda_CH}).
In Fig.~\ref{fig:CH}(\textbf{e}), the value of $\lambda_{v}$ is given for each Voronoi cell as a function of the radial distance $\rho = r - R_0$, $R_0$ being the radius of the coarsening region. For $\rho<0$, $\lambda_{v}$ decreases with the radial position and is broadly distributed, which is a signature of the coarsening dynamics of earlier generated spots. 
Conversely, for $\rho>0$, $\lambda_{v}$ increases linearly with $r$ and is less broadly distributed. This shows that $\lambda$, the typical distance between new spots, scales linearly with the radius $r$. Hence, $\Gamma(r) = \lambda/L$ is actually a constant $\Gamma = \Gamma_0$, which explains why the number of parastichies does not change during the growth of the pattern. 
%\FB{this is very nice!}

Eventually, Fig.~\ref{fig:CH}\textbf{f} shows the data of Fig.~\ref{fig:CH}\textbf{e} rescaled by $\sqrt{\sigma}$. The collapse of all curves is consistent with the results of Fig.~\sref{sfig:CH_scale}\textbf{b}, i.e. what matters is only the order number of the rings of new appearing spots and the number of spots per ring.
To confirm this scenario, Fig.\sref{sfig:CH_R_vs_n}\textbf{a} plots the front radius $R(t)$ at different times as a function of the number of enclosed spots $n(t)$ for different values of $\sigma$.
If $R$ is rescaled by $\sqrt{\sigma}$ (Fig.~\sref{sfig:CH_R_vs_n}\textbf{b}), all data collapse on the same curve which means that, if time is measured in terms of the number $n$ of spots and space by unit of the instability wavelength $\lambda_{CH}$, all the patterns are equivalent (see discussion in \ref{ssec:ssimilar_scaling}). 
Note that, despite this general self-similarity, both highly ordered spiralling patterns (as the one shown in Fig.~\ref{fig:CH}\textbf{a}) and less ordered arrangements (like the one shown Fig.~\ref{fig:CH}\textbf{b} and Fig.~\ref{fig:CH}\textbf{c}) can be observed. This behavior is analysed in detail in the Methods and in figures \sref{sfig:CH_T_conv}, \sref{sfig:CH_noise} and \sref{sfig:CH_S_conv}.

Finally, we observe that the product distribution in reactive fronts is controlled by the initial value of $b$, and the imposed values of $a$ and $Q$ at the inlet, which in our case also dictates the morphology of the expressed pattern via the source term $a b$ in eq.~\ref{eq:phi_prec}.
In particular, the observation of spots instead of stripes 
\cite{Thiele2019}, requires the phase field $\phi$ to be close to the limit of linearly unstable region, i.e. $\int{ab(r,t)dt}\gtrsim\phi_m$. 
The observation of phyllotactic patterns thus requires a fine tuning of the interplay between the reaction--diffusion--advection front and the CH spatial symmetry breaking instability.  

%_____________________________________FIGURE_______________________________________
%_________________________________________________________________________________
\begin{figure*}[!t]
\includegraphics[width=\textwidth]{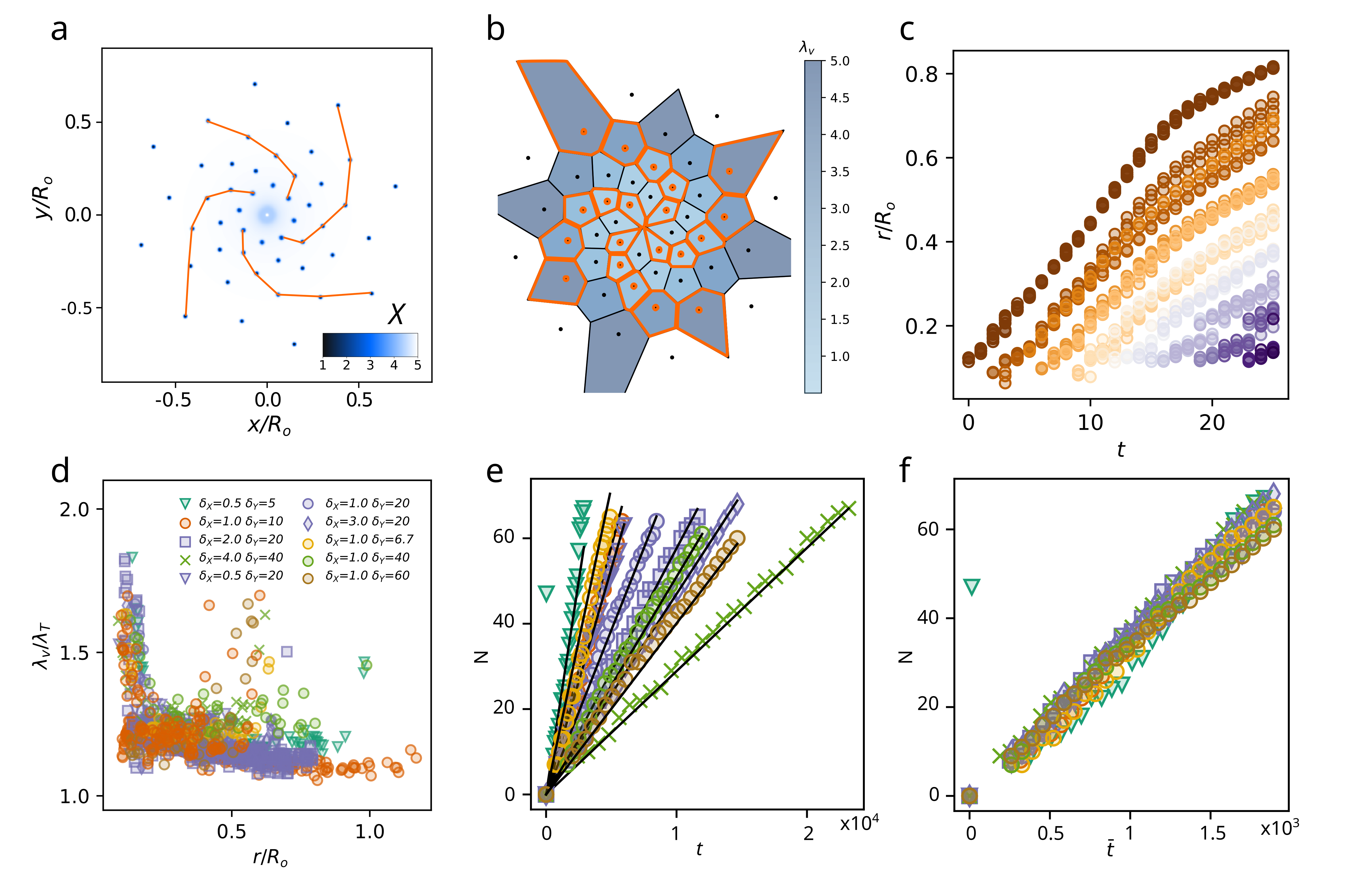}
 \caption{\textbf{Numerical solutions of the Brusselator model and pattern analysis.} \textbf{a:} Two-dimensional distribution of the concentration $X$ in a Turing regime, solution of Eqs.(4)  for $Q=0$, $\delta_B=\delta_X=1$ and $\delta_Y=10$. Initially, ($A,B,X,Y$)=(0,2,0,0) everywhere while $A$ is maintained at $A_0 = 1$ and $B_0=0$ at the circular inlet area of size $R_i = \lambda_0/10$, and no-flux boundary conditions are applied to the other variables. 
Spatial coordinates $x$ and $y$ have been rescaled by $R_o$, the radius of the simulation domain, where $R_o=20\lambda_0$ and 
%$\lambda_0=\lambda_T(r=0,t=0)$
$\lambda_0$ is the characteristic wavelength at the inlet.
 \textbf{b:} Voronoi diagram of the pattern shown in \textbf{a}. Voronoi cells are coloured according to the value of the Voronoi radius $\lambda_{v}$. 
%Note that at larger radius Voronoi cells are larger coherently with the increase of the mutual distance between spots.
Orange contours highlight the spiralling arrangement of the spots.
\textbf{c:} Temporal evolution of the spots radial position for the simulation shown in \textbf{a}. 
Different colors correspond to different spots with earlier spots in brown and later spots in blue.
The radial coordinate $r$ is rescaled by $R_o$.  
%and $R$ is the size of the simulation domain.
Far from the domain edge ($r/R_o = 1$), spots velocity is approximately constant, while this trend is lost approaching the domain boundary.
\textbf{d:} Ratio between the distance $\lambda_{v}$ between spots and the Turing wavelength $\lambda_T$ as a function of their radial position for different simulations.
Points collapse on the same horizontal line $\lambda_{v}/\lambda_T\sim 1.15$ which indicates that the local value of the Turing length controls the distance between points.
Different symbols correspond to different values of $\delta_X$ and different colors correspond to different values of $\delta_Y$.
%\GS{I don't find their values anywhere.}.
\textbf{e:} Total number $N$ of spots as a function of time for simulations with variable values $\delta_X$, $\delta_Y$ and $Q=0$. 
%\ADW{See fig \ref{sfig:BR_diapo}f ???} of the simulation domain. 
New spots appear at constant rate for all the simulations we performed.
\textbf{f:} Same as \textbf{e} but with time $t$ rescaled as $\bar{t}=t/\sqrt{\delta_X\delta_Y}$. All points collapse on the same master curve, suggesting that the point's drift velocity or, equivalently, their appearance rate, is controlled by $\sqrt{\delta_X\delta_Y}$.
}
%WORD COUNT 070623 : 280
\label{fig:BRU}
\end{figure*}

%==================================
\section{Reaction--diffusion Turing patterns}
\label{turing}
Phyllotactic pattern can also be obtained in another class of spatial symmetry breaking instability: Turing patterns. 
To show this, we consider 
%a paradigmatic model for studying temporal oscillations and pattern formation in nonlinear out-of-equilibrium systems 
the irreversible Brusselator model, $A {\to} X$, $ B + X {\to} Y + D$, $2X + Y {\to} 3X$, $ X {\to} E$,  where $\{A, B\}$ are the initial reactants and $\{X, Y\}$ are the autocatalytic $X$ and the inhibitor $Y$ intermediate species with diffusivities $D_x$ and $D_y$, respectively  ~\cite{Prigogine1968}. In extended systems with spatially homogeneous and constant reactant concentrations $A_0$ and $B_0$ (pool chemical approximation \cite{Nicolis1977}), the coupling of the Brusselator kinetics to diffusion can produce spatial stationary Turing structures, characterized by an intrinsic wavenumber: 
\begin{equation}
\label{eq:lambda_bru}
k^2_{T}=\frac{A}{\sqrt{D_X D_Y}},
\end{equation} 
when $B_0>B^T_c=(1+A_0 \sqrt{\delta})^2$, where $\delta = D_X/D_Y< \delta_c=(\sqrt{1+A^2}-1)^2/A^2$. 
The morphology of the stationary Turing structures depends on the value of the control parameter $B_0$, with hexagonally-distributed spots being the first symmetry to appear subcritically followed by supercritical stripes \cite{Dewel1995,DeWit1999}. 

Here, we consider a heterogeneous distribution of the reactants $A(r)$ and $B(r)$ resulting from a radial diffusion or diffusion--advection transport of A from a circular inlet of radius $R_i$, where $A=A_0$ and $B=0$, into a two-dimensional sea where the reactants concentration is initially $A=0$ and $B=B_0$. 
Neglecting the chemical consumption of A and B, the set of dimensionless equations describing such $A+B\rightarrow$ \textit{oscillator fronts} reads~\cite{Nicolis1977,Budroni2016,Budroni2017}:
\begin{subequations}
\begin{align} 
\label{eq:BR_A}
D_t A &= \nabla^2 A, \\
\label{eq:BR_B}
D_t B &= \delta_B\, \nabla^2 B,\\
\label{eq:BR_X}
D_t X &= \delta_X\, \nabla^2 X + A - (B+1)X + X^2 Y, \\
\label{eq:BR_Y}
D_t Y &= \delta_Y\, \nabla^2 Y + BX - X^2 Y,
\end{align}
\end{subequations} 
where $\delta_I=D_I/D_A$ are the dimensionless diffusion coefficients (see also Methods section). 
%and  and the imposed
%The radial advection, $\mathbf{v} = (Q/r) \hat{\mathbf{e}}_r$, results from the imposed constant volumetric flow rate $Q$ at which reactant $A$ is injected at the center of the reactor.
Fig.\ref{fig:BRU}\textbf{a} shows a typical Turing pattern obtained by solving numerically Eqs.(\ref{eq:BR_A}--\ref{eq:BR_Y}) with $Q=0$.
%with radially changing reactant concentration profiles.
Spots of high concentration of the reaction intermediate $X$ arrange themselves along eight distinct spirals as shown in the Voronoi diagram of Fig~\ref{fig:BRU}\textbf{b}, which reminds of phyllotactic patterns.   
Contrarily to experimental precipitations and numerical CH patterns described above, new spots appear here close to the central injection, and successively drift away in the radial direction as shown in Fig.~\sref{sfig:BR_traj} (SI) and \href{https://figshare.com/s/7fdfe30c9f542aa3a3a1}{movie S3}, even in the absence of advection (i.e. $Q=0$). 
The distance between spots increases with the distance from the center as seen on Fig.~\ref{fig:BRU}\textbf{b}.
New spots appear at a constant rate (Fig.~\ref{fig:BRU}\textbf{e}) and their drifting velocity is approximately constant as long as the spots are far from the edge of the simulation domain (Fig.~\ref{fig:BRU}\textbf{c}).
The spot dynamics is therefore directly comparable to the phyllotactic movement of the magnetic droplets in Douady and Couder experiments \cite{Douady1992} even though, in the Turing patterns, the release rate of the new spots and their drift velocity cannot be directly imposed.
Instead, their dynamics is here  a spontaneous consequence of how spots self-organise in space with an intrinsic local wavenumber dependent on the $A(r)$ profile as in eq.~\ref{eq:lambda_bru}.

%MARCELLO: In other terms, these phyllotactic patterns follow the same construction shown in Ref.  \cite{Douady1996b}, with the principle of energy minimization among new and previously generated seeds intrinsically played by the characteristic wavelength of the symmetry breaking instability, which grows radially according to spatially varying reactant profiles.}
Phyllotactic Turing patterns were also observed for other sets of parameters as summarised in Table~\sref{stab:Bru_simulations} and Fig.~\sref{sfig:BR_diapo} and Fig.~\sref{sfig:BR_Q_var} (SI).
By varying the values of $\delta_X$ and $\delta_Y$, we found that the spot dynamics is entirely controlled by the transport of reactant $A$ via the local value of the Turing wavelength $\lambda_T=2 \pi/k_T$, also in the presence of advection $Q$ (i.e. $Q\neq0$).

To prove it, we show in Fig~\ref{fig:BRU}\textbf{d} the ratio between the local distance $\lambda_v(r)$ between spots  (computed as in the previous section) as a function of the radial position and $\lambda_T(r)$ and
observe that all points of each simulation collapse on the same horizontal line at $\lambda_{v}(r)/\lambda_T(r)\sim 1.15$.
Finally, to prove  that the appearance rate $f$ of new spots % is constant, and 
is controlled by the value of $\delta_X$ and $\delta_Y$, we plot in Fig.~\ref{fig:BRU}\textbf{f} the number of spots $N$ as a function of a rescaled time $\bar{t}=t/\sqrt{\delta_X\delta_Y}$. For all simulations performed (see Table~\sref{stab:Bru_simulations}), the data collapse on the same straight line. 
%\MB{Should we mention that this scaling is somehow expected from eq. \ref{eq:lambda_bru}?}.

Note that, as for the CH simulations, the dimensionless number $\Gamma=\lambda/L$ remains constant for the Brusselator patterns,  
coherently with a number of spirals remaining also constant.
However, differently from CH simulations, here $\Gamma$ stays constant because new spots appear in the center, on a circle of approximately constant radius $r=r^*$ as shown in Fig.~\sref{sfig:BR_traj}\textbf{b}, jointly to the negligible variation of $\lambda_T(r^*)$ as shown in Fig.~\sref{sfig:1point_lambda}.
%Our observations indicates that this is the consequence of an initial splitting ad rearrangement of newly formed spots that self-organise at the vertex of an octagon before starting to drift radially as it is visible in movies S??. 

%=========================================

\section{Phyllotactic spatial symmetry-breaking patterns}
Despite the differences in their dynamics and in the nature of the spatial symmetry-breaking instability producing the pattern, phyllotactic structures observed in the phase separation CH model and in the reaction--diffusion Brusselator model show important similarities.
First, in both cases, the pattern selection is tightly controlled by the instability wavelength $\lambda$, through the dimensionless number $\Gamma=\lambda/L$.
In particular, we find that $L$ is 
%either a constant (Brusselator), 
always a function of $\lambda$, i.e. $\Gamma=\Gamma(\lambda)$. In all scenarios explored here, $\Gamma(\lambda)$ turns actually to be a constant but we expect that other scenarios may exist where $\Gamma$ is not constant.

Another strong similarity is that, in both models and in the precipitation experiments as well, the observed patterns belong to the family of the "whorled modes", i.e. phyllotactic structures where the number of left and right turning spirals is the same.
Interestingly, while much less abundant in extant plants as compared to Fibonacci spiral modes, whorled modes may have not been the exception in ancestral plants \cite{Turner2023} and are observed next to spiral modes \cite{Schoute1922,Douady1996b}.
Moreover, accordingly to \cite{Douady1996b}: 1) whorled modes always develop by the appearance of multiple new primordia at once and 2) the selection of whorled or spiral modes is driven by a criterion of packing optimisation.
In the present work, we confirm that both 1) and 2) are retrieved. 
Indeed, groups of new spots appear simultaneously on circular fronts in all our experiments and simulations. %of our phase separation (CH) and Brusselator model.
Also we showed, by using Voronoi diagrams, that phyllotactic patterns observed in CH and Brusselator models arrange themselves on a hexagonal lattice, which corresponds to the most efficient packing solution in 2D. 
%\GS{The above paragraphs just answered many of my criticisms, grats! Sounds lovely.}

If our results appear consistent with previous explorations, 
%\GS{\sout{a} only one } Only one sounds to me a bit arrogant ;-)
one question remains, which is: can one observe spiral modes and Fibonacci transitions in phyllotactic patterns driven by a symmetry-breaking instability in radial geometries, such as those presented here? 
According to the general approach proposed by Gol\'e et al. \cite{Gole2016}, the selection between whorled (or quasi-symmetric) and spiral modes is controlled by the initial noise and the rate of change of $\Gamma$. 
Unfortunately, this hypothesis could not be tested in the minimal model systems presented here because $\Gamma$ was always constant, which means that whorled modes are attractors \textbf{\cite{Gole2016}}. %\MB{if I understand the logic of this sentence, instead of ". However ..", I would write ", and, according to Gol\'e et al., for this case and large enough noise, whorled modes represent attractors, i.e. the stable modes towards which simulations converge after many iterations, which may explain our results."}
%However, as proposed by Gol\'e et al., whorled modes could be attractors, i.e. the stable modes towards which simulations converge after many iterations, in the case of constant $\Gamma$ and large enough noise, which may explain our results.

More generally, we remark that the development of a spatial symmetry-breaking instability is inevitably frustrated in a spatial domain whose typical length is smaller than the instability's intrinsic wavelength.
As a consequence, the corresponding pattern cannot appear but across a region that spans at least some wavelengths.
Thus, the simultaneous appearance of new primordia may be intrinsic to phyllotactic patterns driven by symmetry-breaking instabilities and restrict the access to modes other than whorled modes, which is coherent with previous literature \cite[and citations therein]{Douady1996b}.

To conclude, we have shown that phyllotactic patterns can appear when a spatial symmetry-breaking instability is coupled to the dynamics of a radially growing reaction front. Specifically, spots self-organized along symmetric ontogenic spirals have been obtained experimentally in the wake of a chemical precipitation front and in models of both phase transition and Turing self-organisation when one reactant controlling the dynamics of the reaction front is radially transported into the other one. 
The genericity of the systems analysed, suggests that the observed spiraled patterns 
belong to a new class of phyllotactic growth structures, that we encourage to explore  in the vast class of spatial symmetry-breaking instabilities as observed in a wide range of fields from chemistry \cite{Castets1990} to optics \cite{Haudin2014} and biology \cite{Nakamasu2009, Budrene1995}.

% \end{itemize}

\vskip 5truemm
\section{Methods}
\label{num-methods}
{\small
\noindent\textbf{Analysis of experimental patterns}
Experimental patterns were analysed using the Python library SciPy.
For each experimental image, we manually identify a circular path, centered in the inlet, where the luminosity signal shows a good contrast.
Next, we select all the pixels (highlighted in yellow in Figs.~\ref{fig:experiments}\textbf{a--c}) which belong to an annular region which is 4 pixels thick and is centered on the circular path.
We then report the luminosity $I$ of each pixel as a function of their azimuthal position $\eta$ and perform a moving average of $I(\eta)$.
Next, we perform a linear 1D interpolation of $I(\eta)$ in order to have a regularly sampled signal (see Figs.~\ref{fig:experiments}\textbf{d--f}) and perform a Fast Fourier transform (Figs.~\ref{fig:experiments}\textbf{g--i}).
Finally, we manually identify the maximum of the FFT signal and use the maximum position to estimate the angular distance $\eta_{max}$ between two successive spirals.

\noindent \textbf{Cahn-Hilliard model.} Equations (\ref{eq:A_prec}-\ref{eq:phi_prec}) were solved in two steps.
First, we solve the reaction--diffusion--advection equation (\ref{eq:A_prec}-\ref{eq:B_prec}) in a 1D axisymmetric high resolution domain [$R_i$,$R_o$] using the finite elements solver COMSOL Multiphysics and store the value of the source term $ab$ as a function of $r$ and $t$.
For the reactant concentrations $a$ and $b$, boundary conditions were no-flux at the outer boundary $r=R_o$ and Dirichlet at the inner boundary $r=R_i$ where we fix $a(R_i)=8$ and $b(R_i)=0$. The velocity field was considered as stationary and axisymmetric, i.e. $\boldsymbol{v}(\boldsymbol{r,t})=v(r)$, where $v(r)$ is simply $Q/r$ because the flow is supposed laminar and incompressible.

Next, we solve the CH equation (\ref{eq:phi_prec}) using a homemade finite differences code written in Python in a 2D square grid domain of side $2R_o$.
The value of $ab$ was computed for each grid point and time step by using the linear interpolation function provided by the Python library SciPy and no-flux boundary conditions were imposed at the outer boundaries.
Our investigations showed that the observed pattern is extremely sensitive to the initial conditions which can strongly affect the degree of "ordering" of the final spiralling patterns as mentioned in the results section.
This feature emerged when conducting convergence tests on our simulations. 
In particular, while time convergence is well verified (see Fig.~\sref{sfig:CH_T_conv}), i.e. identical patterns are obtained by taking half the time-step, we face a more complex scenario for spatial convergence.
Indeed, in certain runs it is sufficient to change the gridstep to observe a significant rearrangement of spirals and even the disruption of part of them, which corresponds to patterns similar to those presented in Figs.~\ref{fig:CH}\textbf{b} and \ref{fig:CH}\textbf{c}, while the overall scaling of the pattern is always preserved (Figs.~\ref{fig:CH}\textbf{e} and \ref{fig:CH}\textbf{f}).

\noindent
\textbf{Convergence study} In order to investigate this behaviour and rule out the possibility that our simulations were not well resolved, we proceeded in the following way (see SI section \ref{ssec:numerical_convergence} for more details).
First, we add some random noise to the source term $ab$ and repeat the same simulation as in Figs.~\ref{fig:CH}\textbf{a-b} by only changing the noise seed as shown in Fig.~\sref{sfig:CH_noise}.
Second, we initialize some simulations with the pattern taken at short time from different simulations with the same parameters (including noise) but at different spatial resolution as shown in Fig.~\sref{sfig:CH_S_conv}.
We observed that: 
\begin{itemize}
\item low amplitude noise, i.e. $10^{-3}$ times smaller than the source amplitude, is sufficient to modify the final spiralling pattern.
\item When initiating simulations using the pattern obtained from other simulations at early but sufficiently large time, i.e. 1/10 of the total simulation time, spatial convergence is recovered. 
\end{itemize}  
These two results indicate that, for a given set of parameters, multiple (genuine) solutions are possible and that their selection, as well as
the degree of spirals ordering, is extremely sensitive to multiple factor including initial noise, spatial resolution and source amplitude.

\noindent \textbf{Brusselator model.} 
The set of dimensionless equations (\ref{eq:BR_A}-\ref{eq:BR_Y}) is obtained using the reaction constants $k_i$ as defined in the system of equations \ref{seq:brusselator_reactions}.
The time scale was defined as $\tau=1/k_4$, space scale is $l=\sqrt{\tau D_A}=\sqrt{D_A/k_4}$, $\delta_I=D_I/D_A$ are the dimensionless diffusion coefficient and concentrations were rescaled as it follows: $A=\sqrt{k_1^2 k_3 / k_4^3}[A]$;  $B=\left(k_2 / k_4\right)[B]$; $X=\sqrt{k_3 / k_4}[X]$; $Y=\sqrt{k_3 / k_4}[Y]$; where $[I]$ are the dimensional concentrations. 
Finally, we neglected the reactive consumption of A and B, supposing that their depletion occurs on a much slower time scale as compared to that characterizing the dynamics of the chemical intermediates, as it happens, for instance, in the well-known Belousov–Zhabotinsky reaction.

Equations (\ref{eq:BR_A}-\ref{eq:BR_Y}) were solved using the finite elements solver COMSOL Multiphysics.
The simulation domain has the same circular shape as in CH simulations except in one simulation in a linear geometry.
Space was discretized using a triangular mesh of variable size, and the maximum length of a grid element was set to 1/20 of the minimum wavelength observed in the pattern. 
For species A and B, boundary conditions were no-flux at the outer boundary $r=R_o$ and Dirichlet at the inner boundary $r=R_i$ where $A(R_i)=8$ and  $B(R_i)=X(R_i)=Y(R_i)=0$. 
When the injection flow rate $Q$ is non zero, the velocity field is solved in the same way as in CH simulations. 
}

%TC:ignore 
%----------------------------------------------------------------------------------------
%	REFERENCE LIST
%----------------------------------------------------------------------------------------
\sloppy
%\setcounter{biburllcpenalty}{9000}
%% NOTE I need to replace manually @book with @incollection in the bib file to have decent layout
\bibliographystyle{naturemag}
\bibliography{phyllo}

%TC:endignore 
\end{document}